\documentclass[12pt]{article}
\usepackage{epsf,latexsym,rotate}
\usepackage[ps,arc,frame]{xy}
\usepackage{amsfonts,amssymb} %%amsmath
\epsfverbosetrue
\usepackage[ps,arc,frame]{xy}
\usepackage{graphicx}
\usepackage{slashed}

\newcommand{\lapprox}{%
\mathrel{%
\setbox0=\hbox{$<$}
\raise0.6ex\copy0\kern-\wd0
\lower0.65ex\hbox{$\sim$}
}}
\newcommand{\gapprox}{%
\mathrel{%
\setbox0=\hbox{$>$}
\raise0.6ex\copy0\kern-\wd0   
\lower0.65ex\hbox{$\sim$}
}}

\newcommand{\double}[1]{\mathbb{#1}}

\newcommand{\cc}{\double{C}}

\newcommand{\aaa}{\mathcal{A}}

\newcommand{\hhh}{\double{H}}
\newcommand{\mm}{\mathcal{M}}

\newcommand{\pp}{\pmatrix}

\newcommand{\dd}{\mathcal{D}}

\newcommand{\de}{\hbox{\rm{d}}}

\newcommand{\ot}{\otimes}
\newcommand{\op}{\oplus}

\newcommand{\bb}{\begin{eqnarray}}
\newcommand{\ee}{\end{eqnarray}}
\newcommand{\eee}{\nonumber\end{eqnarray}}
\newcommand{\qq}{\quad}

\newcommand{\rxyh}[2]{{\begin{xy} 0;<2mm,0mm>:<0mm,2mm>::0;0,
,(5,-2)*{a}
,(10,-2)*{b}
,(15,-1.8)*{\bar{b}}
,(20,-2)*{c}
,(25,-1.8)*{d}
,(30,-1.8)*{\bar{d}}
,(2,-5)*{a}
,(2,-10)*{b}
,(1.8,-15)*{\bar{b}}
,(2,-20)*{c}
,(1.8,-25)*{d}
,(1.8,-30)*{\bar{d}}
,(5,-5)*\cir(#1,0){}
,(10,-5)*\cir(#1,0){}
,(15,-5)*\cir(#1,0){}
,(20,-5)*\cir(#1,0){}
,(25,-5)*\cir(#1,0){}
,(30,-5)*\cir(#1,0){}
,(5,-10)*\cir(#1,0){}
,(10,-10)*\cir(#1,0){}
,(15,-10)*\cir(#1,0){}
,(20,-10)*\cir(#1,0){}
,(25,-10)*\cir(#1,0){}
,(30,-10)*\cir(#1,0){}
,(5,-15)*\cir(#1,0){}
,(10,-15)*\cir(#1,0){}
,(15,-15)*\cir(#1,0){}
,(20,-15)*\cir(#1,0){}
,(25,-15)*\cir(#1,0){}
,(30,-15)*\cir(#1,0){}
,(5,-20)*\cir(#1,0){}
,(10,-20)*\cir(#1,0){}
,(15,-20)*\cir(#1,0){}
,(20,-20)*\cir(#1,0){}
,(25,-20)*\cir(#1,0){}
,(30,-20)*\cir(#1,0){}
,(5,-25)*\cir(#1,0){}
,(10,-25)*\cir(#1,0){}
,(15,-25)*\cir(#1,0){}
,(20,-25)*\cir(#1,0){}
,(25,-25)*\cir(#1,0){}
,(30,-25)*\cir(#1,0){}
,(5,-30)*\cir(#1,0){}
,(10,-30)*\cir(#1,0){}
,(15,-30)*\cir(#1,0){}
,(20,-30)*\cir(#1,0){}
,(25,-30)*\cir(#1,0){}
,(30,-30)*\cir(#1,0){}
#2\end{xy}}}

\begin{document}

\font\twelve=cmbx10 at 13pt
\font\eightrm=cmr8

\thispagestyle{empty}

\begin{center}

CENTRE DE PHYSIQUE TH\'EORIQUE $^1$ \\ CNRS--Luminy, Case
907\\ 13288 Marseille Cedex 9\\ FRANCE\\

\vspace{2cm}

{\Large\textbf{Gauge unification in noncommutative geometry}} \\

\vspace{1.5cm}

{\large Christoph A. Stephan $^2$}

\vspace{2cm}

{\large\textbf{Abstract}}
\end{center}
Gauge unification is widely considered to be a desirable feature for extensions
of the standard model. Unfortunately the standard model itself does not 
exhibit a unification of its running gauge couplings but it is required by grand unified theories as well as the noncommutative version of the standard model \cite{cc}.

We will consider here the extension of the noncommutative 
standard model by vector doublets as proposed in \cite{vector}. Two
consequences of this modification are: 1. the relations of the coupling
constants at unification energy are altered with respect to the
well known relation from grand unified theories. 2. The extended model
allows for unification of the gauge couplings
at $\Lambda \sim 10^{13}$ GeV. 
\vspace{2cm}

\noindent
PACS-92: 11.15 Gauge field theories\\
MSC-91: 81T13 Yang-Mills and other gauge theories

\vskip 1truecm

%\noindent CPT-Pxx-2007\\
\noindent \\

\vspace{1.5cm}
\noindent $^1$ Unit\'e Mixte de Recherche  (UMR 6207)
du CNRS  et des Universit\'es Aix--Marseille 1 et 2 et  Sud
Toulon--Var, Laboratoire affili\'e \`a la FRUMAM (FR 2291)\\
$^2$ also at Universit\'e Aix--Marseille 1,
christophstephan@gmx.de\\

It is generally believed that the standard model and the big dessert are not the
final theory describing the particle content of our universe. A hint for an underlying,
more profound structure is the observation that the running gauge couplings almost
converge, missing each other by roughly five orders of magnitude between
$\sim 10^{12}$ GeV and $\sim 10^{17}$ GeV. Grand unified theories require
an exact convergence, but since the standard model cannot provide for this,
extensions have to be considered. One of the most popular extensions
is certainly supersymmetry which enlarges the particle contend of the standard
model roughly by a factor of two, introducing supersymmetric partners. 
Due to cancellations in renormalisation this extension leads to an 
exact convergence of the gauge couplings. The price which has to be
paid is a multitude of hitherto unobserved particles which should although be
detectable at the LHC.

A different approach to the standard model is noncommutative geometry
\cite{book} which, through the spectral action, also requires gauge unification 
\cite{cc,mc2}. Here again the pure standard model cannot meet the conditions
on the gauge couplings. The conditions on the gauge couplings coming from
noncommutative geometry coincide for the standard model with the classical
ones  from grand unified theories. In noncommutative geometry this 
unification is not thought
of as having its origin in the breaking of a simple unifying group 
like $SU(5)$ or $SO(10)$ but as a modification of space-time itself.

Recently extensions of the standard model within the framework of
noncommutative geometry have been discovered \cite{chris,colour,vector}.
At least one of these extensions, the $AC$-model, even has a viable dark matter
candidate \cite{klop} and is compatible with high precision measurements
in particle physics \cite{knecht}.

In this publication we will examine the extension presented in \cite{vector},
investigating its ability to cure the unification problem. Here the particle content
of the standard model is enlarged by particles coupling vectorially to
the electro-weak $U(1)_Y \times SU(2)_w$ subgroup.  A most interesting fact
of these extensions is that the conditions of the gauge unification get modified. 

If the mass of these vector doublets is taken to be of unification scale,
$\sim 10^{13}$ GeV, the altered unification conditions are almost
exactly fulfilled. And even if one prefers the classical conditions from 
grand unified theories, these vector doublets alter the running of the
gauge couplings sufficiently to obtain a perfect convergence.

\section{Vector doublets}

In noncommutative geometry the gauge group $G$ is extracted from the spectral 
triple either via the unimodularity condition \cite{real,mc2} or via centrally
extending the lift of the automorphism group of the associated algebra  \cite{lift}. 
The two  approaches coincide for a minimal
central extension \cite{lift}.

There are other constraints, on the fermionic representations, coming
 from the axioms of the spectral triple. They are conveniently
captured in Krajewski diagrams which classify all possible finite  dimensional spectral triples \cite{kps}. They do for spectral triples  what the Dynkin and weight diagrams do for groups and  representations.

The model considered here is an extension of the  standard model by a set 
of fermions which couple vectorially to the $U(1)_Y \times SU(2)_w$ subgroup
of the standard model. They are colour singlets and have gauge invariant masses
$m_\psi$.
For convenience we will  call them vector doublets. A thorough presentation
of this model containing the details of the construction of the spectral triple, 
the lift of the automorphisms, the Lagrangian
and possible mass assignments, which could give viable dark matter candidates,
can be found in \cite{vector}. 
Extensions of the standard model within the noncommutative framework are
rare and only a few viable ones are known \cite{chris,colour,vector}. Therefore
the vector doublet model is far from ad-hoc and its properties are quite remarkable. 
We will concentrate here on the ability of the model
to achieve unification of the $U(1)_Y$-, $SU(2)_w$- and $SU(3)_c$-gauge couplings.

Figure 1 shows the Krajewski diagram of the standard  model in Lorentzian signature with one generation of fermions and one vector doublet represented by the dashed
arrow. 

\begin{center}
\begin{tabular}{c}
\rxyh{0.7}{
,(5,-20)*\cir(0.3,0){}*\frm{*}
,(5,-25)*\cir(0.3,0){}*\frm{*}
,(5,-20);(10,-20)**\dir{-}?(.4)*\dir{<}
,(5,-20);(15,-20)**\crv{(10,-17)}?(.4)*\dir{<}
,(5,-25);(15,-25)**\crv{(10,-28)}?(.4)*\dir{<}
,(15,-5);(25,-5)**\crv{~*=<2pt>{.}(20,-7.5)}?(.55)*\dir{>}
} \\ \\
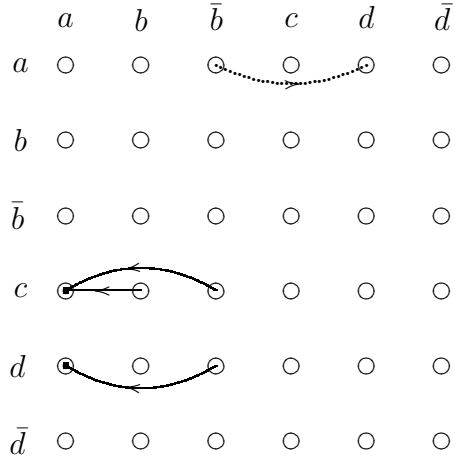
Figure 1: Krajewski diagram for the particle part of the standard model \\
\hskip-1.1cm and the vector doublets depicted by the dashed arrow.
\\
\end{tabular}
\end{center}

We only present the basic components of the finite part of the spectral triple since
they differ from the  standard model. The algebra has four  summands:
$\aaa=\hhh\op\cc\op M_3(\cc)\op\cc\owns (a,b,c,d),$
the Hilbert space  carries the faithful repesentation
$ \rho(a,b,c,d):=
\rho_{L}\oplus\rho_{R}\oplus{\bar\rho^c_{L}}\oplus{\bar\rho^c_{R}}$
with
\bb
\rho_{L}(a,d):=
a\ot 1_3\oplus a \oplus  d 1_2 ,\ 
\rho_{R}(b,d):=b  1_3\oplus\bar b  1_3\oplus b\oplus d \oplus \bar b 1_2, 
\nonumber\\[1mm] 
\rho^c_{L}(a,c,d):=
1_2\ot c\oplus \bar d1_2 \oplus a ,\ 
\rho^c_{R}(a,c,d) :=
c\oplus c\oplus \bar d\oplus\bar d \oplus a.  
\ee
For a detailed treatment of the standard model with four summands in the algebra
we refer to \cite{class}.
The Dirac operator reads
\bb  \dd=\pp{0&\mm&0&0\cr
\mm^*&0&0&0\cr
0&0&0&\bar\mm\cr
0&0&\bar\mm^*&0},
\ee
where  $\mm$
contains the Dirac masses
\bb\mm=
\left[\pp{M_u&0\cr 0&0} \ot 1_3+
\pp{0&0\cr 0&M_d}\ot 1_3\right]
\oplus\left[\pp{M_\nu&0\cr 0&0}+
\pp{0&0\cr 0&M_e}\right] \oplus M_v
\ee
with $M_v$ for the gauge invariant mass matrix of the vector doublets.

For this model all the axioms of noncommutative geometry  \cite{book} are 
fulfilled. Majorana neutrinos may be introduced at the expense of altering
the orientability axiom \cite{ko6}. Note also that this model is  free of gauge
anomalies and 
mixed gauge and gravitational anomalies
for any number of vector doublets. This includes Witten's $SU(2)$ anomaly.
It is also interesting that this models resembles the Connes-Lott model
\cite{lott} regarding the four summands in the algebra.

The  hyper-charge and 
the weak iso-spin of the vector doublet are summarised in table \ref{chargem}.
They follow immediately from the central charges of the standard model
and the requirement that the resulting lift should be minimal, i.e. as less
multi-valued as possible \cite{lift}.

\begin{table}
\begin{center}
\begin{tabular}{|c|c|c|c|c|}
\hline
&I & $I_3$ & $Y_{vec}$ & $Q_{el}$  \\ 
\hline &&&&\\
$(\psi_{1})_{L,R} = \psi^-_{L,R}$ & $2$ & $-\frac{1}{2}$ & $-\frac{1}{2}$ & $-1$  \\
&&&& \\
\hline
&&&& \\
$(\psi_{2})_{L,R} = \psi^0_{L,R}$ & $2$ & $+\frac{1}{2}$ &$-\frac{1}{2}$ & $0$ \\
&&&& \\
\hline
\end{tabular}
\caption{Charge assignment for a negatively charged component}
\label{chargem}
\end{center}
\end{table}

Here $(\psi_{1/2})_{L,R}$ denote the first/second component of the left-handed
or right-handed vector doublet.
Note that after symmetry breaking one component of the vector doublet 
acquires an electric charge while the other becomes electrically neutral. 
This results in a slight mass difference of $\sim 350$ MeV due to radiative
corrections, where the neutral particle is lighter than its charged partner.

\section{The constraints on the gauge couplings}

The spectral action is defined as the number of eigenvalues of the  Dirac operator
up to a cut-off $\Lambda$. As input one has this cut-off, the parameters of the
inner Dirac operator, i.e. fermion masses and mixing angles and three
positive parameters for the cut-off function. 

As an output one obtains the Yang-Mills-Higgs action, in case of the  spectral
triple of the  standard model it is exactly the desired standard model action
\cite{cc,mc2}, and additional constraints on the dimensionless couplings. 
For the standard model with three generations this implies the following relation 
for the gauge  couplings at the cut-off $\Lambda$:
\bb 
5 \,g_1^2=  3\, g_2^2= 3 \, g_3^2, 
\label{4con}
\ee
where $g_1$ is the $U(1)_Y$ coupling, $g_2$ the $SU(2)_w$ coupling and
$g_3$ the  $SU(3)_c$ coupling.
These relations coincide with the unification conditions of grand unified theories. It is
well known that this constraint cannot be met at any unification
scale $\Lambda$ within the standard model
alone and therefore extensions of the standard model have to  be considered.

If we extend the standard model by vector doublets the spectral action
produces a slightly different constraint for the gauge couplings at the
cut-off \cite{vector}:
\bb 
\left(5 + \frac{N_v}{2} \right)  \,g_1^2=  \left( 3\,  + \frac{N_v}{2} \right)  g_2^2= 3 
\, g_3^2, 
\label{constvec}
\ee
where $N_v$ denotes the number of vector doublets.
This is quite remarkable since we have for the first time a deviation from the
classical unification condition (\ref{4con}). 

In  grand unified theories one believes that a simple unifying gauge group
is the reason for the constraints (\ref{4con}).  
In the noncommutative approach we  believe that at the energy $\Lambda$ the noncommutative character of  space-time ceases to be negligible. The ensuing uncertainty relation  in space-time might cure the short distance divergencies and thereby  stabilize the constraints. Indeed Grosse \& Wulkenhaar have an  example of a scalar field theory on a noncommutative space-time whose  $\beta$-function vanishes to all orders \cite{raimar}. It is not  too surprising that additional particles change the constraint (\ref{4con})
since in noncommutative geometry adding new particles means changing the
spectral triple and therefore the geometry itself.

The strategy is now the following. Since the mass of the vector doublets
is gauge invariant it can be  chosen freely. We will choose it in such a way
that the running  couplings  of the standard model plus 
vector doublets meet  conditions (\ref{constvec}) at a given energy scale 
$\Lambda$ which is then identified with the cut-off scale. Furthermore
we will repeat this analysis for the classical conditions (\ref{4con}).

Adding vector doublets changes of course the $\beta$-functions for the
gauge couplings needed to evolve the constraints (\ref{4con}) and (\ref{constvec}).
We restrict ourselves to the one-loop $\beta$-functions.
We set:
$ t:=\ln (E/m_Z),\qq \de g/\de t=:\beta _g,\qq \kappa :=(4\pi )^{-2} $ and we 
will neglect all fermion masses below the top mass and also  neglect threshold effects.

By the Appelquist-Carazzone decoupling theorem  we distinguish two energy domains: $E>m_\psi$ and $E<m_\psi$, where $m_\psi$ is the mass of the vector doublets. For
simplicity we take all vector doublets to have the same mass. 
At high energies, $E>m_\psi$, the $\beta$-functions are for the standard
model with three generations plus $N_v$ vector doublets \cite{mv,jones}:

\bb 
\beta _{g_i}&=&\kappa b_ig_i^3,\qq 
b_i=
{\textstyle
\left( \frac{41}{6}  + \frac{2}{3}\,N_v,-\frac{19}{6}+\frac{2}{3}\,N_v,
-7  \right)  },
\ee
At low energies, $E<m_M$,
the $\beta$-functions are the same with $N_v$ put to zero.
We suppose that all couplings (other than $g_\nu$ and $k$) are  continuous  at 
$E=m_\psi$, no threshold effects.
The three gauge couplings  have  identical evolutions in both energy domains:
\bb g_i(t)=g_{i0}/\sqrt{1-2\kappa b_ig_{i0}^2t}.\ee
The initial conditions are taken from experiment \cite{data}:
$ g_{10}= 0.3575,\ 
g_{20}=0.6514,\ 
g_{30}=1.221.$

Let us first consider constraint (\ref{constvec}). It is not possible to 
obtain an exact unification but by adding one vector doublet, i.e. $N_v=1$,
we obtain a close match for the unification condition of the three gauge couplings
for a vector doublet mass of $m_\psi = 4\times 10^{13}$ GeV,
see table \ref{gauge1}.  The slight mismatch  could be explained with the
transition to a truly noncommutative space-time.
\begin{table}
\begin{center}
\begin{tabular}{|c|c|c|c|}
\hline
&&&\\
&$\left(5 + \frac{1}{2} \right)  \,g_1^2=   \left( 3 + \frac{1}{2} \right) g_2^2$ &
$\left(5 + \frac{1}{2} \right)  \,g_1^2=  3\, g_3^2 $& 
$\left(3 + \frac{1}{2} \right)  \,g_2^2=  3\, g_3^2  $ \\ 
&&&\\
\hline &&&\\
$\Lambda $ & $8,8\times 10^{13}$ GeV & $2,8 \times 10^{13}$ GeV & 
$6,6 \times10^{12}$ GeV  \\
&&&\\
\hline
\end{tabular}
\caption{Unification energies  for $N_v=1$ with condition (\ref{constvec})}
and $m_\psi = 4 \times 10^{13}$ GeV
\label{gauge1}
\end{center}
\end{table}
Since the mass of the vector doublets is gauge invariant  it is natural to 
take it of the order of $\Lambda$.

If we add more vector doublets we do no ameliorate the situation as the
example for $N_v=2$ with $m_\psi = 4\times 10^{13}$ GeV shows, see table
\ref{gauge2}. 
\begin{table}
\begin{center}
\begin{tabular}{|c|c|c|c|}
\hline
&&&\\
&$\left(5 + \frac{2}{2} \right)  \,g_1^2=   \left( 3 + \frac{2}{2} \right) g_2^2$ &
$\left(5 + \frac{2}{2} \right)  \,g_1^2=  3\, g_3^2 $& 
$\left(3 + \frac{2}{2} \right)  \,g_2^2=  3\, g_3^2  $ \\ 
&&&\\
\hline &&&\\
$\Lambda $ & $5,5\times 10^{14}$ GeV & $4,7 \times 10^{12}$ GeV & 
$4,6 \times10^{10}$ GeV  \\
&&&\\
\hline
\end{tabular}
\caption{Unification energies  for $N_v=2$ with condition (\ref{constvec})}
and $m_\psi = 4 \times 10^{13}$ GeV
\label{gauge2}
\end{center}
\end{table}
This is due to the effect of the  vector doublets on the unification condition 
(\ref{constvec}).
Their influence on the actual running of the coupling is rather small because 
their mass is comparable to the cut-off scale. Smaller masses for the vector
doublets also spoil the ability to meet condition (\ref{constvec}). 

To complete our analysis let us now repeat the preceding considerations, but
with respect to the  classical unification condition (\ref{4con}). What we want to
show is how easy it is to achieve gauge unification with a minimal extension
of the standard model. The masses of the vector doublets are in this classical
setting much lower. We have summarised the results for $N_v=2$ and 
$N_v=3$ in table \ref{gauge3}.  Note that the unification condition (\ref{4con})
can be met exactly.
$N_v=1$ results in a vector doublet mass
below 10 GeV and should be experimentally excluded.
\begin{table}
\begin{center}
\begin{tabular}{|c|c|c|}
\hline
&&\\
$N_v$ &$m_\psi$ & $\Lambda $  \\ 
&&\\
\hline &&\\
$2$ & $1,2\times 10^{4}$ GeV & $5 \times 10^{13}$ GeV\\
&&\\
\hline &&\\
$3$ & $2,0\times 10^{7}$ GeV & $5 \times 10^{13}$ GeV\\
&&\\
\hline
\end{tabular}
\caption{Unification energies and vector doublet masses  for $N_v=2,3$ with condition (\ref{4con})}
\label{gauge3}
\end{center}
\end{table}

\section{Conclusions}

Noncommutative geometry as well as grand unified theories impose constraints 
on the gauge couplings of Yang-Mills-Higgs models. They are assumed to
be valid at a certain energy scale $\Lambda$, the unification scale.
For the standard model these constraints coincide in the
noncommutative setting and in the grand unified setting. But, since these
conditions cannot be fulfilled when taking into account only the standard 
model particle content one assumes that the big dessert has to be
populated.

We analysed here an extension of the standard model  by vector doublets 
within noncommutative
geometry \cite{vector}. This extension exhibits two main features:

\begin{itemize}
\item adding the vector doublets changes the constraint that the
gauge couplings have to fulfil at unification scale
\item they allow for gauge unification at $\sim 10^{13}$ GeV 
with respect to the new set of constraints and with respect to the classical constraints
from grand unification
\end{itemize} 

The masses of the vector doublets are in the  case of the modified 
constraints of the order of the unification scale. In the classical case
they range from $1,2 \times 10^4$ GeV to $2,0 \times 10^7$ GeV depending
on the number of doublets added.

It is certainly possible to build more baroque models from the extensions
proposed in \cite{chris,colour} and \cite{vector} which also allow for gauge
unification. But the model examined here has certainly the appeal of being
very minimal.

\vskip1cm
\noindent
{\bf Acknowledgements:} The author would like to thank T. Sch\"ucker
for helpful comments and discussions and gratefully acknowledges a fellowship of the Alexander von Humboldt-Stiftung.

\end{document}